\documentclass[fleqn, 12pt, a4paper, twoside]{article}
\usepackage{parco2005}

\usepackage{fancyhdr}
\usepackage{graphicx}
\usepackage{fancyvrb}
\usepackage{pslatex}

\newtheorem{theorem}{Theorem}

\begin{document}
%\maketitle

%\textwidth 41pc

\title{Asynchronous iterative computations with Web information retrieval
structures: The PageRank case}
\author{
Giorgos Kollias\address{Computer Engineering and Informatics Department,
University of Patras, GREECE},
Efstratios Gallopoulos\addressmark,
Daniel B.\ Szyld\address{Department of Mathematics,
Temple University, Philadelphia, USA}
}

%\bibliographystyle{abbrev}

%\begin{document}

\maketitle

\section{Introduction \label{sec:intro}}
There are several ideas being used today for Web information retrieval, and specifically in Web search engines
\cite{langville:survey:2005}. The PageRank algorithm \cite{page:pagerank:1998} is one of those that introduce a content-neutral
ranking function over Web pages. This ranking is applied to the set of pages returned by the Google search engine in response
to posting a search query. PageRank is based in part on two simple common sense concepts:
%\begin{itemize}
%\item
%\\
(i)
A page is important if many important pages include links to it.
%\item
%\\
(ii)
A page containing many links has reduced impact on
the importance of the pages it links to.
%\end{itemize}

In this paper we focus on asynchronous iterative schemes \cite{bertsekas-tsitsiklis:book:1989,frommer:asynchronous:2000} to
compute PageRank over large sets of Web pages. The elimination of the synchronizing phases is expected to be advantageous on
heterogeneous platforms. The motivation for a possible move to such large scale distributed platforms lies in the size of
matrices representing Web structure. In orders of magnitude: $10^{10}$ pages with $10^{11}$ nonzero elements and $10^{12}$
bytes just to store a small percentage of the Web (the already crawled); distributed memory machines are necessary for such
computations. The present research is part of our general objective, to explore the potential of asynchronous computational
models as an underlying framework for very large scale computations over the Grid \cite{Grid.2}. The area of ``internet
algorithmics'' appears to offer many occasions for computations of unprecedent dimensionality that would be good candidates for
this framework.

After giving a formulation of PageRank and its common interpretations in Section \ref{sec:2}, we present its treatment under
synchronous computational models. We next consider the asynchronous approach and comment on key aspects, specifically
convergence, termination detection and implementation.
% These sections contain both references to related work and elements of
% our perspective as implemented next.
In Section \ref{sec:exp}, we describe the experimental framework and present preliminary numerical experiments, while in
Section \ref{sec:concl} we draw our conclusions and discuss our future work on this topic. In this paper, as is common
practice, we do not address the effects of finite precision arithmetic and roundoff error.

\section{Formulation and Interpretations \label{sec:2}}
In order to appreciate the PageRank computation,
we present its standard formulation using the following set
of four $n \times n$ matrices, where $n$ is the number of pages
being modeled.

An \emph{adjacency matrix $A$} can be obtained through a web crawl or
synthetically generated using statistical results, e.g., as in
\cite{broder:graph:2000}.
Thus,
$A_{ij}=1$ \emph{iff} page $i$ points to page $j$, and $A_{ij}=0$ otherwise.

A \emph{transition matrix $P$} has nonzero elements $P_{ij}  = {A_{ij}}/{{\rm deg}(i)}$ when ${\rm deg}(i)\neq 0$, and zero
otherwise (in which case page $i$ is called a \emph{dangling} page); here ${\rm deg}(i) = \sum_{j}{A_{ij}}$ is the outdegree of
page $i$.

A \emph{stochastic matrix $S$} is given by $S = P^{T} + w\;d^{T}$; $w = \frac{1}{n} e$, where $e$ is the size $n$
vector of all
1's,
% so
% that $w$ i.e., a uniform vector normalized to unity,
and $d$ is the \emph{dangling index vector} whose
   nonzero elements are $d_{i} = 1$ \emph{iff} ${\rm deg}(i) = 0$.

The \emph{Google matrix $G$} is  $G = \alpha \;S + (1-\alpha) \; v \;e^{T}$. For a random web surfer about to visit his next
page, the relaxation parameter $\alpha$ is the probability of choosing a link-accessible page. In choosing otherwise, i.e.,
with probability $1-\alpha$, from the complete Web page set vector $v$ contains respective conditional probabilities of such
\emph{teleportations}. Typically $v = w$ and $\alpha = 0.85$.

The PageRank vector $x$ is the solution of the linear system
\begin{equation}
  \label{eq:pagerank-fixed-point}
  x = G \; x ~,
\end{equation}
where the matrix $G$ is an irreducible stochastic matrix, and thus its largest eigenvalue in magnitude is $\lambda_{max}=1$
\cite{stewart:markov:1994}. Thus, the PageRank vector $x$  is the eigenvector corresponding to $\lambda_{max}=1$, and when
normalized, 
it is the reachability probability in a random walk on the Web, i.e.,
the invariant measure or stationary probability distribution
of a Markov process modeled by the matrix $G$. It can also be computed as the solution to a system of linear equations.
Using the fact that $x$ is normalized to unity, i.e., $e^{T}\;x = 1$, equation (\ref{eq:pagerank-fixed-point}) yields
\begin{equation}
  \label{eq:linear-equation-system}
  (I - R) \; x = b
\end{equation}
where $b = (1-\alpha)\; v$ and $R = \alpha\; S$
is the \emph{relaxed stochastic matrix}
\cite{delcorso:LNCS:2004}.
% Although these interpretations are not of direct practical use, apart from
% providing useful alternative insights,
%The above interpretations provide insight into
%alternative methods for the computation of PageRank: There is a vast literature
%with -overlapping in some cases- methods for computing the invariant
%measure of a Markov process
%\cite{stewart:markov:1994},
%finding the largest modulus eigenvector of a
%matrix and solving a system of linear equations
%\cite{golub-loan:book:1983}.

\section{Synchronous PageRank \label{sec:sync}}
To make the computation of $x$ practical for the problem sizes we are considering, it is necessary to employ an iterative
method, e.g., executing until convergence 
\begin{equation}
  \label{eq:iteration-operator}
  x(t+1) \leftarrow f(x(t))
\end{equation}
with  $t = 0,1,\ldots$ for a suitable operator, $f$ and some initial vector
$x(0)$. The vector $x(t)$ denotes the approximation to $x$ obtained after $t$
iterations. The above process needs to be mapped on
 a specific execution environment, corresponding to a computational model that
typically  preserves the semantics of the mathematical model in (\ref{eq:iteration-operator}).
The environment constitutes
 a virtual machine for the computation and is largely characterized by the types of units of
execution (UE) (e.g., processes, threads) and communication mechanisms
(e.g., shared memory, message passing) it readily
supports, especially in hardware. Execution and communication entities
are ultimately hosted by actual machines typically
attached to nets (e.g., clusters) and internets (e.g., the Internet).
% Mapping (\ref{eq:iteration-operator}) to an execution
% environment leads to a computational model. It is often the case that such a model

In the single UE case the aforementioned mapping on the execution environment is straightforward. For multiple UEs, however,
this requires care: In the shared memory case, a semantics preserving mapping must involve synchronized access to shared memory
cells between cooperating UEs, protected by locks, whereas
in the message passing case, this synchronization is achieved
through a barrier mechanism implemented atop collective blocking communication. For the
% \linebreak
 PageRank computation,  we can easily turn
(\ref{eq:pagerank-fixed-point}) into the following simple iteration:
\begin{equation}
  \label{eq:synchronous-power}
  x(t+1) = G\;x(t), \ \ \ x(0) \mbox{\ given}.
\end{equation}
That is, $f(\cdot)$ amounts to a matrix-vector multiplication.
This is the well-known power method for finding the eigenvector of
$G$ corresponding to the eigenvalue of largest magnitude
\cite{stewart:markov:1994},
except that no per-step normalization needs to be performed.
The normalization is not needed since
a stochastic matrix such as $G$ does not alter
$\|x(t)\|_{1}$; and thus no danger of overflow or underflow
is present here.

Single UE implementations of (\ref{eq:synchronous-power})
with an emphasis on convergence acceleration, support for
personalization through different teleportation vectors and utilization of naturally occurring block structure in the adjacency
matrix $A$ can be found in
\cite{haveliwala:personalizing:2003,kamvar:block:2003,kamvar:extrapolation:2003}. For multiple UEs,
message passing computation of PageRank using the formulation
(\ref{eq:linear-equation-system}) was presented in 
\cite{gleich:parallel:2004}.

\section{Asynchronous PageRank \label{sec:async}}
Unfortunately, the necessary per-step synchronization of the synchronous algorithm described above grows into a significant
overhead, especially as it is
 governed by the rate of the slowest UE  and the costs of
lock or barrier management. %, thus introducing idle cycles in faster UEs.
One radical transformation to harness this problem is to reduce the requirement for synchronization, e.g., by using
non-blocking access to shared memory cells or network buffers. A central theme of our work is to investigate the effect of this
transformation on the convergence, speed and overall effectiveness of the computations.

%%The central idea of our work is to study the effect of this transformation.
%
% In order to alleviate synchronization costs in multiple UE environments one can adopt a drastic approach:
% Do not synchronize at all e.g.  Essentially \emph{data
%imports for each UE are not controlled by algorithm demands but by data availability}; if no new data is available, recycle the
%old one, else pick the newest UE-locally available, but not yet consumed data.

For an environment with $p$ UEs,  denote by $x_{\{i\}}$ the set of indices assigned to $i^{th}$ UE during the iterative
computation, $T^{i}$ the set of times at which $x_{\{i\}}$ is updated
(i.e., $i^{th}$ UE finishes its computation) and $\tau_{j}^{i}(t)$ the
time when the fragment $x_{\{j\}}$, which is available at time $t$ in
the $i^{th}$ UE, was actually produced at its respective $j^{th}$
UE. Then for $t\in T^{i}$, the $i^{th}$ UE updates
\begin{equation}
  \label{eq:general-asynchronous-iteration}
x_{\{i\}}(t+1)\leftarrow f_{i}(x_{\{1\}}(\tau_{1}^{i}(t)),\ldots,
x_{\{p\}}(\tau_{p}^{i}(t)) ),
\end{equation}%
while $x_{\{i\}}(t+1) = x_{\{i\}}(t)$ at other times.
Delays due to omission of synchronization phases are expressed as
differences $t - \tau_{j}^{i}(t) \geq 0$.
The relation (\ref{eq:general-asynchronous-iteration})
is the asynchronous analog of
(\ref{eq:iteration-operator}) where $f_i$ expresses the distributed operator
component executing at the $i^{th}$ UE.
Obviously the form of $f_i$ is independent of the asynchronism introduced.
It thus follows that the normalization-free power method for
PageRank computation  at  the $i^{th}$ UE reads
\begin{equation}
  \label{eq:asynchronous-pagerank-power}
  x_{\{i\}}(t+1) = G_{i}\;[ x_{\{1\}}^\top(\tau_{1}^{i}(t)),\ldots,
x_{\{p\}}^T(\tau_{p}^{i}(t))] ^{\top}
\end{equation}
for $t\in T^{i}$, and $x_{\{i\}}(t+1) = x_{\{i\}}(t)$ at other times, where $G_{i}$ is a set of rows of the Google matrix $G$
indexed by $\{i\}$. Alternatively, while the synchronous, linear system equation approach would lead to an iterative scheme of
the form $x(t+1) = R\;x(t) + b$ which can be seen to be identical to (\ref{eq:synchronous-power}), its asynchronous formulation
would lead to another, slightly different computational kernel, namely
\begin{equation}
  \label{eq:asynchronous-pagerank-linear}
  x_{\{i\}}(t+1) = R_{i}\;[ x_{\{1\}}^\top(\tau_{1}^{i}(t)),\ldots,
x_{\{p\}}^T(\tau_{p}^{i}(t))] ^{\top} + b_{i}
\end{equation}
for $t\in T^{i}$,
and $x_{\{i\}}(t+1) = x_{\{i\}}(t)$ at other times,
at the $i^{th}$ UE.
Here $R_{i}$ is a set of rows
of the relaxed stochastic matrix $R$
indexed by $\{i\}$,
 and $b_{i}$ is the corresponding set of elements of
vector $b$.

Also of interest are P2P computations of PageRank
\cite{jager:mscthesis:2004,sankaralingam:p2p:2003,shi:p2p:2003,%
tsimonou:mscthesis:2004} These fall into the multiple UEs, message passing category and are asynchronous in nature. An
important novelty in these studies
 is the dynamically generated link information through a notification protocol proposed to be
integrated with the host Web servers.
%%EG: GIWRGO PLEASE CHECK THAT THIS IS SO

The lack of synchronization annuls the semantics of the original
 mathematical algorithm. Therefore, it becomes necessary to discuss the convergence
 properties of
the asynchronous scheme (\ref{eq:general-asynchronous-iteration}). We discuss this and related issues in the remainder of this
section.
% convergence question as well as comments on the
% protocol chosen for global convergence detection (something far from trivial in an asynchronous setting) and techniques for
% and  implementing asynchronism are discussed next.

\subsection{Convergence \label{sec:conv}}
Convergence of asynchronous iterative algorithms is usually established through constructing a sequence of nested boxed sets in
the spirit of the following theorem \cite{bertsekas-tsitsiklis:book:1989}:
\begin{theorem} \label{box:theo}
Let $\{X(k)\}:\;\ldots \subset X(k+1) \subset X(k) \subset \ldots \subset X $,
with the following two conditions.
%\begin{itemize}
%\item
\\
Synchronous Convergence Condition: For all $k=1, \ldots$,
%$\forall k,\,$
 $x \in X(k), f(x) \in X(k+1)$, and
for $\{y^k\},\, y^k \in X(k):$  the limit points of $\{y^k\}$ are
fixed points of $f$.
%\item
\\
Box Condition: For all $k=1, \ldots$,
%$\forall k:\, $
$X(k) = X_1(k) \times \ldots \times X_p(k)$. \\
%\end{itemize}
Then  if $x(0) \in X(0) $, the limit points of $\{x(t)\}$ are
fixed points of $f$,
where $\{x(t)\}$ are given by (\ref{eq:general-asynchronous-iteration}).
\end{theorem}

%Theorem \ref{box:theo} applies directly to the iterative kernels of .
Process (\ref{eq:asynchronous-pagerank-power}) involves a nonnegative matrix of unit spectral radius; it is proved in
\cite{lubachevsky:convergence:1986} that the corresponding asynchronous iteration converges to the true solution within a
multiplicative factor that can easily be factored out in the end by renormalization. A discussion on the misconception by some
authors that for a nonnegative matrix $B$, spectral radius $\rho(B) < 1$ is a necessary condition for convergence of an
asynchronous normalization-free power method  can be found in \cite{szyld:mystery:1998}. On the other hand, process
(\ref{eq:asynchronous-pagerank-linear}) involves a matrix $R$ with $\rho(R) < 1$. Asynchronous iterations with such matrices
are well known to converge to the true solution \cite{bertsekas-tsitsiklis:book:1989}.

\begin{figure}[htbp]
  \centering
\begin{tabular}{|c|c|}\hline
\textbf{computing UE} & \textbf{monitor UE} \\ \hline
\begin{minipage}{0.4\textwidth}
  \begin{tabbing}
if\=(checkConvergence()) \\
 \>if\=(not converged) \\
 \>\>converged = true \\
 \>pc++ \\
 \>if\=(pc = pcMax) \\
 \>\>send(CONVERGE, monitor) \\
 \>\>recv(STOP, monitor) \\
 else\\
 \>if\=(converged) \\
 \>\>converged = false \\
 \>\>send(DIVERGE, monitor) \\
\>\>pc = 0
\end{tabbing}
\end{minipage}
&
 \begin{minipage}{0.4\textwidth}
\begin{tabbing}
recv(CONVERGE $|$ DIVERGE, all)\\
if\=(checkConvergence()) \\
 \>if\=(not converged) \\
 \>\>converged = true \\
 \>pc++ \\
 \>if\=(pc = pcMax) \\
 \>\>send(STOP, all) \\
 else\\
 \>if\=(converged) \\
 \>\>converged = false \\
 \>\>pc = 0
\end{tabbing}
\end{minipage}
\\ \hline
\end{tabular}
  \Caption{\texttt{pc}: persistence counter, \texttt{pcMax}: its max value;
reaching it triggers \texttt{CONVERGE/STOP} messages.
They can have different values in monitor, computing UEs;
all: all computing UEs}
  \label{fig:termination-detection}
\end{figure}%

\subsection{Termination Detection \label{sec:td}}
The termination
of asynchronous iterative algorithms is a non-trivial matter since
local convergence at an UE does not automatically ensure global
convergence. Even in the extreme case when all UEs have locally
converged, one can devise scenarios where messages not yet delivered could
destroy  local convergence.

Both centralized and distributed protocols for termination detection
can be found in the literature,
\cite{baz:termination:1996,savari:termination:1996}.
In a centralized approach, a special UE acts as a monitor of
the convergence process of other computing UEs; it keeps a log of the
convergence status and issues \texttt{STOP} messages to
all computing UEs when all of them have signaled their local convergence.
In fact, computing UEs can issue either \texttt{CONVERGE}
(when achieving local convergence) or \texttt{DIVERGE}
(when exiting such a state) messages to the monitor UE.
Distributed protocols for global convergence detection
(see, e.g., \cite{tanenbaum-steen:book:2002}) are flexible but
rather complex to implement. They typically
assume a specific underlying communication topology.
For example in \cite{bahi:termination:2005} a leader
election protocol is used, which in turn assumes a tree topology.

Our draft version
of a practical centralized protocol,
in part inspired by \cite{bahi:jace:2004},
is presented in Figure~\ref{fig:termination-detection}. It
enforces \emph{persistence} of
convergence both at the computing UEs (for issuing a
\texttt{CONVERGE} message) and at the monitor UE
(for issuing a \texttt{STOP} message).
Persistence is introduced to provide time
for pending -and perhaps divergence causing- messages to be actually delivered.
%assuming that not listings package is provided...

\subsection{Implementation \label{sec:impl}}
We focus on multiple UEs message passing environments,
which is the case for our experiments.
In that case, we need non-blocking communication primitives.
These are actually implemented either by using multithreading
(e.g., one thread per communication channel)
or by multiplexing  such channels and probing from within  a single
thread (through \texttt{select()} type mechanisms) for new data. Since
multiple messages might have been received in the meantime, messages should be
kept in queues organized under a common discipline.

\section{Numerical Experiments \label{sec:exp}}
\subsection{Application Structure}
Our application consists of scripts steering \texttt{Java} classes.
These scripts are written in \texttt{Jython}~\cite{jython-python:software},
which is an
implementation of the \texttt{Python}~\cite{python:software}
programming language in \texttt{Java}~\cite{java:software}.
Such a mixed-language approach facilitates writing
portable, interactive, easily extensible and flexible systems; after
all, performance critical
operations can always be isolated into compiled \texttt{.class} code.

Scripts build and use objects. \texttt{Configuration} objects can load/store parameters from/to configuration files -
accessible from all other objects, partition and distribute matrix or vector data and optionally send code or launch processes
over the cluster nodes. \texttt{Computation} objects perform computations and exchange information related to convergence status with
\texttt{Monitor} objects implementing the termination detection protocol;
cf.~Figure~\ref{fig:termination-detection}.
Communications are established through
\texttt{Communication} objects which set up suitable communicators upon their instantiation; these communicators expose
communication primitives to be invoked at each step.

We use multithreading in order to implement non-blocking communications. An asynchronous \texttt{send()} or \texttt{recv()} is
just its blocking counterpart wrapped in a thread object and submitted to a thread pool endowed with a suitable task-handling
strategy. Data are imported/exported through read/write channels with locks synchronizing those concurrently executing threads
which happen to be managing messages with identical source and target IDs. Access to thread pool queues and pending
communication-task-handles is provided so that a customized
thread-management policy can be applied. At startup, a single file
containing computation parameters should be available.
This file is used by a \texttt{Configuration} object for the generation
of node-specific configuration files and a script for distributing these files (optionally with other data or updated source
code files) to the cluster nodes and initiating the computation. An option for automatic report generation is also provided.

\subsection{Numerical Results}
We used a Beowulf cluster of Pentium-class machines at 900 MHz, with 256 MB RAM each, running Linux, version 2.4 and connected
to a 10 Mbps Ethernet LAN. We used Java 5.0, Jython 2.1 and \emph{Matrix Toolkits for Java} \cite{mtj-java:software} for
composing our scripts and classes, all freely available on the Web. We report on some of the results of this ongoing work. The
transition matrix used in the experiments is the Stanford-Web matrix
  \cite{stanfordmatrix:data}, generated from an actual web-crawl. It contains connectivity
 info for $281,903$ pages ($2,312,497$ non-zero elements, $172$ dangling
 nodes).
We used  the computational kernel (\ref{eq:asynchronous-pagerank-power})
with a local convergence threshold of $10^{-6}$. Note that in each case, blocks of consecutive $[n/p]$ rows were distributed among computing machines.
Termination detection used \texttt{pcMax = 1} on both monitor and computing
UEs. Configurations with 2, 4, and 6 machines were tested for both synchronous
and asynchronous computations.
\begin{table}[htbp]
  \centering
\begin{tabular}{|c|c|c|c|c|c|}\hline
  \multicolumn{1}{|c|}{} &
  \multicolumn{2}{|c|}{\textbf{Synchronous}} &
  \multicolumn{2}{|c|}{\textbf{Asynchronous}} &
  \multicolumn{1}{|c|}{} \\ \hline
  \textbf{procs} &
  \textbf{$iters$} &
  \textbf{$t$ (sec)} &
  \textbf{$[iters_{min}, iters_{max}]$} &
  \textbf{$[t_{min}, t_{max}]$ (sec)} &
  \textbf{$\langle speedUp \rangle$} \\ \hline
  2 & 44 & 179.2 & [68, 69]& [86.3, 94.5] & 1.98 \\ \hline
  4 & 44 & 331.4 & [82, 111]& [139.2, 153.1] & 2.27 \\ \hline
  6 & 44 & 402.8 & 129, 148]& [141.7, 160.6] & 2.66 \\ \hline
\end{tabular}
  \caption{Numerical results: For the asynchronous case iteration ranges,
  computation time ranges are given ([max, min] values) since local
  convergence threshold is not `simultaneously' reached at all nodes. A column
  with the average speedup offered by asynchronous computation over
  synchronous one is given (averages are over extreme values in the
  asynchronous case).}
\label{tab:timing-results}
\end{table}
Results in Table \ref{tab:timing-results} are encouraging. On the other hand, it is fair to note that
 they correspond to
reaching local convergence threshold. Assembling vector fragments resulting from asynchronous computations at monitor UE and
then checking global convergence reveals that a threshold of the order of $5\times 10^{-5}$ has actually been reached.
Preliminary results of timing with respect to reaching a common global threshold (instead of a local one) reveals a modest
speedup of asynchronous vs. synchronous computation in the $10-20\%$ range. Responsibility for the degradation of performance
when increasing the number of UE's appears to lie with the overall large communication-to-computation ratio of the current
algorithm. Observe, however, that what is important are not the accurate values of the PageRank vector components, but their
relative ranking. Therefore,  an issue in our present investigations is the effect of a more relaxed global threshold criterion
on the computed page ranks.
% A very large
%communication to computation ratio
%  coupled with all-to-all communication is responsible for getting longer
% computations even when going to more UEs configurations; however our comparison
%  strictly concerns synchronous vs.\ asynchronous performance.

Asynchronous iterative algorithms also seem to naturally adapt to heavy communication demands in a computation; current
\texttt{send()/receive()} threads can block but computation thread is free to advance to next step iteration. On the contrary,
in synchronous mode, no option exists except for blocking all threads (even the computation one), until data emitted from all
nodes actually reach their destinations and synchronization completes, no matter whether the supporting network's
characteristics suffice. In this case asynchronous computation can exhibit a low message import ratio (always with respect to
iteration count which is obviously increased relative to synchronous setting); see Table \ref{tab:interaction-results}.
\begin{table}[htbp]
  \centering
\begin{tabular}{|c|c|c|c|c|c|}\hline
  \multicolumn{1}{|c|}{} &
  \multicolumn{4}{|c|}{\textbf{Sender}} &
  \multicolumn{1}{|c|}{} \\ \hline
  \textbf{Receiver} &
  \emph{id = 0} &
  \emph{id = 1} &
  \emph{id = 2} &
  \emph{id = 3} &
  \textbf{Completed Imports (\%)} \\ \hline
  \emph{id = 0} & 109 & 46 & 23 & 26 & 29 \\ \hline
  \emph{id = 1} & 40 & 107 & 22 & 27 & 28 \\ \hline
  \emph{id = 2} & 35 & 37 & 111 & 66 & 41 \\ \hline
  \emph{id = 3} & 27 & 30 & 54 & 82 & 45 \\ \hline
\end{tabular}
  \caption{Completed imports for the 4 computing UEs, asynchronous case. Rows
  contain the number of different vector fragments actually received during the computation from
  peers with respective IDs. Diagonal numerical entries contain the total number of locally
  computed and thus locally used vector fragments. \emph{Completed Imports} column
  contains percentage averages of imports actually completed (should all be
  $100\%$ for the synchronous case).}
  \label{tab:interaction-results}
\end{table}%
\section{Conclusions and Future Work \label{sec:concl}} The major performance
bottleneck in our experiments to date is due to the large volume of data and  the frequency that it is being produced. The
latter is caused by  the small computation time per-iteration (sparse matrix-vector multiplication). Note also that the
communication pattern is an all-to-all scheme at each step; all these factors conspire to surpass the available network
bandwidth and thus build memory consuming buffers of pending messages at the sending ends.

In the case of asynchronous iterations, data is being produced at a rate that is even higher than in the synchronous case,
because part of the time gained from  eliminating the synchronization phases is actually used for the production of extra
messages; these (favorably) advance local iteration counters but they could also (unfortunately) overload the network; we guard
against this misfortune by cancelling \texttt{send()/recv()} threads not having completed within a time window. The following
is thus a hardly surprising conclusion from our experiments: Asynchronous iterative algorithms make up an alternative
computation methodology in distributed environments. However this is not a black-box methodology and is most effectively
utilized by iterative methods with heavy computational component and light communication. A frequent, all-to-all, fat message
passing can saturate network infrastructure capacity, even in modest but dedicated cluster environments; heterogeneous
environments like the Grid would be even more sensitive to such message passing scenarios. We would thus like to avoid the use
of  all-to-all communication schemes; after all the flexibility of
  asynchronous iterations gives us a choice on the targets of produced
  messages. Furthermore, it is advisable to employ an adaptive communication scheme; if message sending/receiving tasks fail to
  complete within a number of local iterations, reduce the
  rate of message exchanges with this not well `responding' node.
%% But even in that case, the
%% flexibility of asynchronous iterative schemes - even without further tuning - could lead to better performance.
%\begin{itemize}
%\item Do not use all-to-all communication schemes; after all the flexibility of
%  asynchronous iterations gives us a choice on the targets of produced
%  messages.
%\item Use an adaptive communication scheme; if message sending/receiving tasks fail to
%  complete within a number of local iterations, reduce the
%  rate of message exchanges with this not well `responding' node.
%\end{itemize}

In our ongoing work,  we explore adaptive schemes for the asynchronous computation of PageRank. We also experiment with
\texttt{select()} based implementations of asynchronism in order to amortize thread management costs. Since trees are naturally
occurring internetwork topologies we also plan to study the performance of moving a clique-based (i.e., all-to-all) synchronous
iterative method to an asynchronous, tree-based counterpart. We are also considering the use of suitable permutations (cf.
\cite{choi:partitioning:1996}) as well as larger data sets.
% We intend to present results from this current research at the
% conference and include them in the final version of this paper.

\paragraph{\bf Acknowledgments}
The work of the first two authors was partially supported by a Hellenic Pythagoras-EPEAEK-II research grant. The work of the
third author was partially supported by the US NSF grant CCF-0514489. The authors would also like to thank Professor G.\ Kallos
for providing access to the computational facilities of the Physics Department, University of Athens.

% \bibliography{gdk}

%% \bibliography{gdk,supplement-parco2005}

\begin{thebibliography}{10}

\bibitem{java:software}
Java language website.
\newblock \texttt{http://java.sun.com}.

\bibitem{jython-python:software}
Jython website.
\newblock \texttt{http://www.jython.org}.

\bibitem{mtj-java:software}
{Matrix Toolkits for Java website}.
\newblock \texttt{http://www.math.uib.no/~bjornoh/mtj/}.

\bibitem{python:software}
Python language website.
\newblock \texttt{http://www.python.org}.

\bibitem{stanfordmatrix:data}
Stanford {W}eb {M}atrix.
\newblock \texttt{http://nlp.stanford.edu/~sdkamvar/data/stanford-web.tar.gz}.

\bibitem{bahi:termination:2005}
J.M. Bahi, S.~Contassot-Vivier, R.~Couturier, and F.~Vernier.
\newblock A {D}ecentralized {C}onvergence {D}etection {A}lgorithm for
  {A}synchronous {P}arallel {I}terative {A}lgorithms.
\newblock {\em IEEE Trans. Parallel Distrib. Syst.}, 16:4--13, 2005.

\bibitem{bahi:jace:2004}
J.M. Bahi, S.~Domas, and K.~Mazouzi.
\newblock Jace: {A} {J}ava {E}nvironment for {D}istributed {A}synchronous
  {I}terative {C}omputations.
\newblock In {\em EUROMICRO-PDP'04}, pages 350--357. IEEE, 2004.

\bibitem{baz:termination:1996}
D.El Baz.
\newblock {A} method of terminating asynchronous iterative algorithms on
                  message passing systems.
\newblock In {\em Parallel {A}lgorithms and {A}pplications}, 9:153--158, 1996.

\bibitem{bertsekas-tsitsiklis:book:1989}
D.P. Bertsekas and J.N. Tsitsiklis.
\newblock {\em Parallel and Distributed Computation}.
\newblock Prentice Hall, Englewood Cliffs, NJ, 1989.

\bibitem{broder:graph:2000}
A.~Broder, R.~Kumar, F.~Maghoul, P.~Raghavan, S.~Rajagopalan, R.~Stata,
  A.Tomkins, and J.~Wiener.
\newblock Graph structure in the web: experiments and models.
\newblock In {\em 9th Int'l. WWW Conf.}, 2000.

\bibitem{choi:partitioning:1996}
H.~Choi and D.B.\ Szyld.
\newblock Application of threshold partitioning of sparse matrices to {M}arkov
  chains.
\newblock In {\em IPDS'96}, pages 158--165. IEEE, 1996.

\bibitem{jager:mscthesis:2004}
D.~{de Jager}.
\newblock Page{R}ank: {T}hree {D}istributed {A}lgorithms.
\newblock Master's thesis, Imperial College of Science, Technology and
  Medicine, London, Sept. 2004.

\bibitem{delcorso:LNCS:2004}
G.M. {Del Corso}, A.~Gulli, and F.~Romani.
\newblock Fast {PageRank} computation via a sparse linear system.
\newblock In {\em Lecture Notes in Computer Science, Vol.~3243}, pages
  118--130. 2004.

\bibitem{Grid.2}
I.~Foster and C.~Kesselman, editors.
\newblock {\em The Grid: Blueprint for a New Computing Infrastructure}.
\newblock Morgan Kaufmann - Elsevier, San Francisco, 2004.

\bibitem{frommer:asynchronous:2000}
A.~Frommer and D.B. Szyld.
\newblock On asynchronous iterations.
\newblock {\em J. Comput. Appl. Math.}, 123:201--216, 2000.

\bibitem{gleich:parallel:2004}
D.~Gleich, L.~Zhukov, and P.~Berkhin.
\newblock Fast {P}arallel {P}age{R}ank: {A} {L}inear {S}ystem {A}pproach.
\newblock Technical report, {Yahoo! Inc.}, 2004.

\bibitem{haveliwala:personalizing:2003}
T.H. Haveliwala, S.D. Kamvar, and G.~Jeh.
\newblock An {A}nalytical {C}omparison of {A}pproaches to {P}ersonalizing
  {P}agerank.
\newblock Technical report, Stanford Univ., July 2003.

\bibitem{kamvar:block:2003}
S.D. Kamvar, T.H. Haveliwala, C.~D. Manning, and G.H. Golub.
\newblock Exploiting the {B}lock {S}tructure of the {W}eb for {C}omputing
  {P}age{R}ank.
\newblock Technical report, Stanford Univ., March 2003.

\bibitem{kamvar:extrapolation:2003}
S.D. Kamvar, T.H. Haveliwala, C.~D. Manning, and G.H. Golub.
\newblock Extrapolation {M}ethods for {A}ccelerating {P}age{R}ank
  {C}omputations.
\newblock In {\em Proc. 12th Int'l. WWW Conf.}, May 2003.

\bibitem{langville:survey:2005}
A.N. Langville and C.D. Meyer.
\newblock A {S}urvey of {E}igenvector {M}ethods for {W}eb {I}nformation
  {R}etrieval.
\newblock {\em {SIAM} Rev.}, 47:135--161, 2005.

\bibitem{lubachevsky:convergence:1986}
B.~Lubachevsky and D.~Mitra.
\newblock {A Chaotic, Asynhronous Algorithm for Computing the Fixed Point of a
  Nonnegative Matrix of Unit Spectral Radius}.
\newblock {\em J. ACM}, 33:130--150, Jan. 1986.

\bibitem{page:pagerank:1998}
L.~Page, S.~Brin, R.~Montwani, and T.~Winograd.
\newblock The {P}age{R}ank {C}itation {R}anking: {B}ringing {O}rder to the
  {W}eb.
\newblock Technical report, Stanford Univ., 1998.

\bibitem{sankaralingam:p2p:2003}
K.~Sankaralingam, S.~Sethumadhavan, and J.~C. Browne.
\newblock Distributed {P}agerank for {P2P} {S}ystems.
\newblock In {\em 12th Int'l. Symposium on High Performance Distributed
  Computing}, 2003.

\bibitem{savari:termination:1996}
S.A. Savari and D.P. Bertsekas.
\newblock Finite termination of asynchronous iterative algorithms.
\newblock {\em Parallel Computing}, 22(1):39--56, 1996.

\bibitem{shi:p2p:2003}
S.-M. Shi, J.~Yu, G.~Yang, and D.~Wang.
\newblock Distributed {P}age {R}anking in {S}tructured {P2P} {N}etworks.
\newblock In {\em ICPP'03}. IEEE, 2003.

\bibitem{stewart:markov:1994}
W.J. Stewart.
\newblock {\em Introduction to the Numerical Solution of Markov Chains}.
\newblock Princeton Univ. Press, 1994.

\bibitem{szyld:mystery:1998}
D.B. Szyld.
\newblock The mystery of asynchronous iterations convergence when the spectral
  radius is one.
\newblock Technical Report 98-102, Department of Mathematics, Temple Univ.,
  Philadelphia, Oct. 1998.

\bibitem{tanenbaum-steen:book:2002}
A.S. Tanenbaum and M.~{van Steen}.
\newblock {\em Distributed Systems, Principles and Paradigms}.
\newblock Prentice-{H}all, Upper Saddle River, NJ, 2002.

\bibitem{tsimonou:mscthesis:2004}
L.~Tsimonou.
\newblock {Distributed PageRank: Comparisons between a Simulation and a
  Peer-to-Peer Implementation of the Algorithm}.
\newblock Master's thesis, Imperial College of Science, Technology and
  Medicine, London, Sept. 2004.

\end{thebibliography}
%\endgroup

\end{document}